\documentclass[twocolumn]{article}

\usepackage{amsmath}
\usepackage{amsmath}
\usepackage{amssymb}
\usepackage{physics}

\usepackage{blindtext}
\usepackage{bookman}

\usepackage[T1]{fontenc} 
\usepackage{microtype} 

\usepackage[english]{babel} 

\usepackage[hmarginratio=1:1,top=32mm,columnsep=20pt]{geometry}
\usepackage[hang, small,labelfont=bf,up,textfont=rm,up]{caption} 
\usepackage{booktabs}
\usepackage{lettrine} 

\usepackage{enumitem} 
\setlist[itemize]{noitemsep} 

\usepackage{abstract} 

\usepackage{titlesec} 
\titleformat{\section}[block]{\large\scshape\centering}{\thesection.}{1em}{} 
\titleformat{\subsection}[block]{\large}{\thesubsection.}{1em}{} 
\titleformat{\subsubsection}[block]{\it}{\thesubsubsection.}{1em}{}

\usepackage{fancyhdr} 
\usepackage{titling} 

\usepackage{hyperref} 
\hypersetup{
    colorlinks=true,
    linkcolor=blue,
    filecolor=magenta,      
    urlcolor=blue,
    citecolor=blue,
    pdfpagemode=FullScreen,
    }

\usepackage[dvipsnames]{xcolor}
\usepackage[affil-it]{authblk}
\usepackage{amssymb, physics, slashed}
\usepackage{graphicx}

\usepackage{physics}

\title{Electroweak $\eta_{\rm w}$ meson}
\author{
\textsc{Gia Dvali$~^1$, Archil Kobakhidze$~^2$ and Otari Sakhelashvili$~^2$} 
\vspace{0.2cm} \\
\normalsize \itshape
$^1~$Arnold Sommerfeld Center, Ludwig-Maximilians-Universität,\\ \normalsize \itshape
Theresienstraße 37, 80333 München, Germany and \\ \normalsize \itshape
Max-Planck-Institut für Physik, Boltzmannstr. 8, 85748 Garching, Germany
\vspace{0.2cm} \\ 
\normalsize \itshape
$^{2}~$Sydney Consortium for Particle Physics and Cosmology, \\
\normalsize  \itshape
School of Physics, The University of Sydney, NSW 2006, Australia \\ 
}
\date{} 
\renewcommand{\maketitlehookd}{

\begin{abstract}
\noindent 

We argue that the Standard Model is accompanied by a new pseudo-scalar degree of freedom, $\eta_{\rm w}$-meson, which cancels the topological susceptibility of the electroweak vacuum and gets its mass from this effect. The prediction is based on the analyticity properties of the Chern-Simons correlator combined with the basic features of gravity. Depending on the quality level of the $U(1)_{B+L}$-symmetry, $\eta_{\rm w}$ emerges as a  $B+L$ pseudo-Goldstone boson or as a St\"uckelberg $2$-form of the electroweak gauge redundancy. An intriguing scenario of the first category is the emergence of $\eta_{\rm w}$ in the form of the phase of a $U(1)_{B+L}$-violating fermion condensate triggered by the instantons, somewhat similar to $\eta'$-meson in QCD. Regardless of its origin, the presence of $\eta_{\rm w}$-meson in the theory appears to be a matter of consistency.

\end{abstract}
}

\begin{document}
\maketitle

\section{Introduction} 

In this paper, we argue that the Standard Model is accompanied by a new degree of freedom, the $\eta_{\rm w}$-meson, which gets its mass from the topological susceptibility 
 of the vacuum (TSV) of the electroweak theory. 

{ The main evidence for $\eta_{\rm w}$ comes from noticing 
 that elimination of the electroweak $\theta$-vacuum by the anomalous $B+L$-symmetry of the Standard Model,
 necessarily implies the generation of the massive pole in 
 electroweak TSV. Moreover, this pole has the quantum numbers 
 that fully match the phase of the fermion condensate 
 generated by the electroweak instantons.}
 
The precise nature of the  $\eta_{\rm w}$-meson depends on the quality of the  $U(1)_{B+L}$-symmetry, which is defined as follows.  We shall say that the quality of the $U(1)_{B+L}$-symmetry is good if it is explicitly broken exclusively by the electroweak instantons. In the opposite case, we shall say that the quality of the $U(1)_{B+L}$-symmetry is poor.  This would be the case if, for example, the  $U(1)_{B+L}$-symmetry were to be explicitly broken by some high-dimensional fermion operators generated by physics beyond the Standard Model.

{ We must stress that the fact of existence of the $\eta_{\rm w}$-meson in the spectrum is insensitive to the quality 
of $U(1)_{B+L}$-symmetry. However, its precise nature 
can be. }

{ In arriving at the above conclusions, gravity plays an important role of a monitor.  At the most conservative level, the role of gravity is in guaranteeing the impossibility of the decoupling of fields at finite Planck mass, $M_P$.
Since this property is a direct consequence of general covariance, we take it as common knowledge rather than an extra assumption. The conclusion about the existence 
of the $\eta_{\rm w}$-particle is reached when the above feature of gravity is superimposed over the following correspondence
\cite{Dvali:2005an}: \\

 \noindent {\it  At finite coupling, the nullification of TSV is equivalent to a Higgs phase of the Chern-Simons $3$-form.}  \\

The above correspondence is rather general and follows from the gauge invariance and the analytic properties of the TSV correlator. 

Now, since TSV is nullified by the anomalous $B+L$-symmetry of the Standard Model, the existence of $\eta_{\rm w}$ is inevitable since the emergence of a massive pole is the only option for 
vanishing TSV. 

   We show that a possible explicit breaking of $B+L$ by high-dimensional operators coming from physics beyond the Standard Model (BSM) cannot change the above conclusion. 
   
Next, we move beyond the basic framework.  
By imposing further restrictions, such as the incompatibility of TSV with the requirement of a well-defined $S$-matrix for gravity 
\cite{Dvali:2018dce, Dvali:2022fdv, Dvali:2023llt}, we gain more information about the nature of $\eta_{\rm w}$.
The latter requirement, although fully justified, 
 goes beyond the common knowledge category. 
Using it as a guideline, in case of a poor-quality $U(1)_{B+L}$-symmetry, it motivates to introduce $\eta_{\rm w}$ as a $2$-form $B_{\mu\nu}$ that transforms under the electroweak $SU(2)$ gauge redundancy \cite{Dvali:2005an}. 
 }

 Concerning the most conservative case, an intriguing question, 
to which we devote a special discussion, 
is whether the $\eta_{\rm w}$-meson could emerge as a phase of a fermion condensate triggered by the electroweak instantons. For illustrative purposes in sec. \ref{cond_ch}, we explicitly compute the $(B+L)-$violating condensate within a simplified toy model that carries some relevant features of the electroweak sector of the Standard Model. If the condensate indeed forms,  $\eta_{\rm w}$ would play the role somewhat analogous to $\eta'$-meson of QCD in the limit of a massless up-quark. The mass of the latter is generated from the TSV of QCD. 
Of course, an important difference is that,  unlike QCD,  $SU(2)_w$ is in the Higgs phase. This triggers a number of open questions about the validity domain and the role of the condensate that we shall discuss. 
 
 In summary, regardless of its origin, the $\eta_{\rm w}$-meson 
 appears to be a crucial ingredient for a consistent coupling 
 between the Standard Model and gravity.

\section{Evidence for \texorpdfstring{$\eta_{\rm w}$}{Lg} from topological susceptibility}
\label{2}

\subsection{General argument} 

    In this chapter, we shall deduce the inevitability of the 
  $\eta_{\rm w}$-boson from TSV. 
 Before discussing the $U(1)_{B+L}$-symmetry of the Standard Model, we shall give general arguments 
 connecting the elimination of the $\theta$-vacuum 
 with the existence of a pseudo-scalar.  
 We shall follow  \cite{Dvali:2005an, Dvali:2005ws,  Dvali:2013cpa, Dvali:2016uhn, Dvali:2017mpy, Dvali:2022fdv, Dvali:2023llt}, 
 relying on  gauge redundancy and analyticity
  properties of the spectral representation of TSV.

  We start by formulating the physics of the $\theta$-vacuum 
 in the language of a topological susceptibility.   
      Let us first consider an $SU(N)$ gauge theory 
      with $\theta$-term included in the action,  
       \begin{eqnarray} \label{ST} 
  S_{\theta}\, &=&  \int_{3+1}\, {\theta} F\tilde{F}\,, \nonumber  \\
  &&\, F \tilde{F} \equiv \, \, \epsilon^{\mu\nu\alpha\beta}\, 
 {\rm tr} \, F_{\mu\nu}F_{\alpha\beta} \,,
\end{eqnarray}
where $F_{\mu\nu}$
 is the standard field-strength 
of the $N\times N$ gluon matrix $A_{\mu} \equiv A_{\mu}^cT^c$,  
with $T^c$ the  generators of $SU(N)$ and $c=1,2,...,N^2-1$
the color adjoint index.  This term is a total derivative,  
  \begin{equation} \label{dCS}
  F \tilde{F} \,  = \, 
   \epsilon^{\mu\nu\alpha\beta} \partial_{\mu} C_{\nu\alpha\beta}^{(\rm{CS})}\,,
   \end{equation}
 where, 
   \begin{equation} \label{CS}
   C_{\mu\nu\alpha}^{(\rm{CS})} \equiv {\rm tr} \left(A_{[\mu}\partial_{\nu}A_{\alpha]} + \frac{2}{3}A_{[\mu}A_{\nu}A_{\alpha]}\right) \,,
   \end{equation}
is the Chern-Simons  $3$-form.
 It thereby can be rewritten as a boundary term, 
   \begin{equation} \label{BoundaryC}
  S_{\theta} \, = \, 
  {\theta} \int_{2+1} dX^{\mu} \wedge dX^{\nu} \wedge  dX^{\alpha}
   C_{\mu\nu\alpha}^{(\rm{CS})} \,,
   \end{equation}
 where $X^{\mu}$ are the embedding coordinates of the 
 $2+1$-dimensional boundary. 
   Of course, the term is invariant since under the 
   gauge transformation, 
 $A_{\mu} \rightarrow U(x)A_{\mu} U^{\dagger}(x) 
+  U \partial_{\mu} U^{\dagger}$ with  
$U(x) \equiv e^{-i\omega(x)^bT^b}$, 
 the $3$-form shifts by an exterior derivative 
 \begin{equation} \label{Cgauge}
C^{\rm CS}_{\mu\nu\beta} \rightarrow C^{\rm CS}_{\mu\nu\beta} + \partial_{[\mu}\Omega_{\nu\beta]} \,,
\end{equation}
where $\Omega_{\mu\nu} = {\rm tr} A_{[\mu}\partial_{\nu]}\omega$. 
  
   First, let us assume that ${\theta}$ is 
a physically observable parameter. In particular, this implies that there exists no anomalous symmetry under which  $\theta$ can be shifted to zero. For example, this is the case in 
    QCD with no massless quarks as well as in the electroweak theory with explicitly broken  $B+L$-symmetry beyond the electroweak anomaly.   As it is well known, in both theories the corresponding $\theta$-terms have observable physical effects. A well-known example of an observable quantity in QCD is the electric dipole moment of the neutron.   
       
     The physicality of the  $\theta$-term  is directly linked      
   with the correlator usually 
     referred to as TSV, 
  \begin{eqnarray}
    \label{EEcorr}
FT \langle F\tilde{F},  F\tilde{F}\rangle_{p\to 0} &&\equiv \\
\equiv \lim\limits_{p \to 0}\int d^4 x e^{ipx} \langle T [F\tilde{F}(x),F\tilde{F}(0)]\rangle &&=\mathrm{const} \nonumber\,,
\end{eqnarray}
where $T$ stands for time-ordering, $FT$ stands for Fourier transformation and  $p$ is a four-momentum.  The $\theta$-term is physical if the above correlator 
is non-zero and vice versa. 
Notice that the correspondence between a non-zero TSV and the existence of 
physical $\theta$-vacua is independent of the specifics of the physics 
that contributes in TSV.  For example, in pure Yang-Mills theories, 
the well-established contributors are the instantons.  However,  for the time being, we shall keep the discussion maximally general.

We now come to the following important point. 
The non-zero value of the expression (\ref{EEcorr})  implies that the 
  K\"all\'en-Lehmann 
   spectral representation of the Chern-Simons correlator includes a physical pole 
    at $p^2=0$ \cite{Dvali:2005an, Dvali:2022fdv, Dvali:2023llt},     
    \begin{equation}
    \label{CCcorr}
FT\langle C^{(\rm{CS})},C^{(\rm{CS})}\rangle =  \frac{\rho(0)}{p^2} + \sum_{m\neq 0} 
\frac{\rho(m^2)}{p^2 - m^2} \,, 
  \end{equation} 
   where $\rho(m^2)$ is a spectral function.   
   The important thing is that $\rho(0) \neq 0$. That is, the physicality of $\theta$ is in one-to-one correspondence 
  with the presence of the pole $p^2 =0$ in   
  (\ref{CCcorr}).

We must stress that, while the correlator in (\ref{CCcorr}) is gauge-dependent, its pole structure is not. The massless 
pole $p^2 =0$ cannot be gauged away, in full analogy with a physical pole in the correlator for a $U(1)$ photon field. 

   The gauge independence of the massless pole in (\ref{CCcorr})
   can be seen in several ways. 
   First, this can be understood by computation of 
   gauge-invariant force mediated by the $3$-form field between the two  conserved sources $J^{\alpha\beta\gamma}(x)$, 
   where $\partial_{\alpha}J^{\alpha\beta\gamma}(x) =0 $. For example,
 \begin{equation}
 J^{\alpha\beta\gamma}(x) = q\int d^3\xi
 \delta(x- X(\xi)) \frac{\partial X^{\alpha} \partial X^{\beta} \partial X^{\gamma}}{\partial \xi^a \partial \xi^b \partial \xi^c} \epsilon^{abc} 
 \end{equation} 
 describes a current of a 
 membrane with $(2+1)$-dimensional world-volume coordinates 
 $\xi^{a},~a=0,1,2$ and $(3+1)$-dimensional target space coordinates $X^{\alpha}, ~\alpha = 0,1,2,3$.
 Here $q$ is a constant representing the $3$-form charge of
 the membrane.  In particular, such a membrane can also 
 describe the boundary given in (\ref{BoundaryC}). For a planar membrane located in the plane 
  $x_3=L$,  the current is 
  $J^{\alpha\beta\gamma}(x) = 
 \delta(x_3 -L) \delta_{0}^{[\alpha} \delta_{1}^{\beta} \delta_{2}^{\gamma]}$. It is easy to see that the correlator evaluated between 
 two parallel static membranes located in the planes 
 $x_3 =0$ and $x_3 = L$ gives a long-range linear potential 
  $V(L) \propto q^2\, L$, which unambiguously indicated that 
  the $3$-form contains a massless field corresponding to
  the $p^2 =0$ pole in the correlator (\ref{CCcorr}). 
  
 The second way of demonstrating the gauge independence of the massless pole is by introducing the St\"uckelberg two-form field 
$B_{\mu\nu}$ for the gauge redundancy (\ref{Cgauge}) as this 
was done in \cite{Dvali:2005an}.   Combined with St\"uckelberg, 
the $3$-form becomes gauge invariant:  
$(\bar{C}^{\rm CS}_{\mu\nu\beta} \equiv C^{\rm CS}_{\mu\nu\beta} \, + \, f \, \partial_{[\mu}B_{\nu\beta]})$, 
where $f$ is a scale. 
Computation of the gauge-invariant version of the 
correlator, $FT\langle \bar{C}^{(\rm{CS})}, \bar{C}^{(\rm{CS})}\rangle$
shows that the pole gets shifted to a non-zero value 
$p^2 = m^2 \propto  1/f^2 $, fully controlled 
by the scale $f$. Thus, taking  $f \rightarrow \infty $,
we gradually recover a massless pole in a fully gauge-invariant manner.  We shall discuss this version of the theory 
in more detail later in connection  with the gauge formulation 
of the axion \cite{Dvali:2005an}.

Thus, in a theory with physical $\theta$-vacuum,  the operator expansion of 
 $C^{(\rm{CS})}$ contains a massless field  $C$.  This is an exact statement. 
Notice that, since $C^{(\rm{CS})}$ is a $3$-form, its massless entry 
brings no propagating degrees of freedom. It is important to emphasize that the understanding of TSV in terms of a massless $3$-form does not amount to any modification of the theory. 
It is just an alternative language for accounting
the $\theta$-vacua.  However, this language  shall allow us to draw powerful conclusions using the equation (\ref{CCcorr}) as a guideline.  
  
  Let us now assume that some new physics added to the theory 
  makes the previously-non-zero TSV  (\ref{EEcorr}) zero. Two different ways of achieving this will be discussed later.  The important point is that regardless of 
 the particular dynamics,  the vanishing TSV implies that  
 the spectral representation (\ref{CCcorr}) no longer contains a massless pole. This can only happen in two ways: 
   1) Either the pole gets shifted to a non-zero value 
   $p^2 = m^2_{\eta} \neq 0$;  or 
   2) The spectral weight of the massless entry 
   vanishes $\rho(0) = 0$. 
   That is, the massless $3$-form field $C$ either becomes massive 
 \cite{Dvali:2005an}  
   or decouples \cite{Hebecker:2019vyf}.  
    
The decoupling option, $\rho(0) = 0$, however, is excluded by gravity. 
The point is that at a finite value of the Planck scale, $M_P$, gravity excludes the existence of any fully decoupled $3$-form field.  Indeed, even if we assume for a moment that, 
 after being canonically normalized, this field is decoupled 
 from all the particle excitations of the gauge sector, it must still couple to gravity.  Due to this, the massless $3$-form has a physical effect.  For example, its field strength contributes to the vacuum energy that sources gravity. Thus, the physical effect of $\theta$-vacua would persists
 gravitationally.  This would be in clear contradiction 
 with the fact that $\theta$ is unphysical because of 
 zero TSV.  
 Thus, the decoupling of the  $p^2=0$ entry is prohibited 
 by gravity at finite $M_P$. 
 
 We are thereby left with the sole option that
 with any deformation of the theory that makes TSV zero, the would-be 
 massless pole shifts to a massive one, $p^2 = m^2_{\eta}$. 
 A massive $3$-form, however, propagates a single degree of freedom and is a pseudo-scalar.  This can be interpreted as $3$-form Higgs effect  \cite{Dvali:2005an}: initially-massless $3$-form that contains no propagating degree of freedom 
 acquires a single degree of freedom and becomes a propagating pseudo-scalar. 
 In conclusion: \\

\noindent {\it  Any physics that eliminates TSV, 
 and thus renders  $\theta$ unphysical,
 leads to the emergence of a massive pseudo-scalar degree of freedom.}

 \subsection{Two ways of removing TSV}
 
  We now discuss the two alternative physics that can 
  ensure the vanishing TSV. As already established by the 
  general argument,  the emergence of 
  a massive pseudo-scalar is inevitable in both cases. 
  However, the nature of this field is different.  
  
\subsubsection{Removing TSV via 
good-quality anomalous \texorpdfstring{$U(1)$}{LG}-symmetry} 
 
   Let us assume that we endow 
   the theory with an anomalous $U(1)$-symmetry
   with the corresponding $U(1)$-current  $J_{\mu}$
    exhibiting an anomalous divergence, 
    \begin{equation}\label{JAnom}
    \partial^{\mu} J_{\mu} \propto  F\tilde{F} \,.
    \end{equation} 
   In such a case the $\theta$-term  (\ref{ST}) can be arbitrarily redefined and in particular, can be set to zero by a proper 
    $U(1)$-transformation,  
      \begin{equation}
      \label{Tshift}      
  \theta \, \rightarrow \, \theta \, +\, const. \,. 
      \end{equation}      
  This implies that the $\theta$-parameter must become unphysical. 
  
Thus, the inclusion of an anomalous symmetry
 must make TSV  (\ref{EEcorr}) zero. 
 As we have already established, this implies a 
 $3$-form Higgs effect and the corresponding emergence of a massive 
 pseudo-scalar \cite{Dvali:2005an}. 
The effect can be viewed as the topological mass generation \cite{Dvali:2005ws} and represents a general consequence of the equations (\ref{EEcorr}) and (\ref{JAnom}).

Now, matching the quantum numbers and the anomaly 
   properties, it is clear that the above degree of freedom 
   must realize the anomalous $U(1)$-symmetry 
   non-linearly.    Hence, in the case of a good-quality 
   anomalous $U(1)$-symmetry, the pseudo-scalar must 
   materialize as a pseudo-Goldstone boson 
   emerging from the spontaneous breaking 
   of the very same $U(1)$-symmetry. 
      
    A useful example illustrating this case is 
    the removal of TSV of QCD by a good-quality
    anomalous symmetry.  
    Such anomalous chiral symmetry in QCD can be obtained 
  either 
by suppressing the Yukawa coupling constant of one of the quarks, or by introducing an   
anomalous Peccei-Quinn symmetry \cite{Peccei:1977hh} via the extension 
 of the field content.  In both cases, the 
 $\theta$-term can be rotated away by the corresponding 
 anomalous $U(1)$-transformation. 
   As a result, the $\theta$-term becomes unphysical. 
   Of course, simultaneously, the TSV vanishes. 
   
   Thereby, as discussed in \cite{Dvali:2005an}, in both 
    cases, we end up with a $3$-form Higgs effect with   
  the required pseudo-Goldstone degree of freedom 
  automatically provided in both realizations of 
  the anomalous $U(1)$-symmetry. In the case of 
  the Peccei-Quinn scenario, this degree of freedom is an axion \cite{Weinberg:1977ma, Wilczek:1977pj}, which comes as a Goldstone boson of the enlarged scalar sector of the theory.   
  Likewise, in the case of a massless quark, the role of the 
  axion is assumed by the $\eta'$-meson of QCD.

 In the latter case, the degree of freedom emerges as the phase 
 of the 't Hooft determinant \cite{tHooft:1976snw}, 
      \begin{equation}
      \label{tDET}
          \mathcal{L}_{'t Hooft} \propto |{\rm det} (\bar{q}_L q_R)| {\rm e}^{i\left(\frac{\eta'}{f_{\eta}} \, - \,{\theta}\right )}  \,.  
      \end{equation}       
where $q_L$ and $q_R$ are left- and right-handed components of $N_f$ flavors of quark fields and 
$f_{\eta}$ is the decay constant. 
   This determinant has a non-zero vacuum expectation value (VEV). Correspondingly, $\eta'$ emerges as 
   a pseudo-Goldstone  boson of the chiral symmetry,       
   \begin{equation}
      \label{Qchiral}      
  q_L \rightarrow {\rm e}^{i\alpha} q_L\,, ~~ q_R  \rightarrow {\rm e}^{-i\alpha} q_R\,.  
      \end{equation}
  This symmetry is broken both spontaneously (via the quark condensate) as well 
  as explicitly (via the chiral anomaly). Due to the anomaly, under the chiral transformation 
  $\theta$ shifts as, 
    \begin{equation}
      \label{TAlpha}      
  \theta \, \rightarrow \, \theta \, +\, 2N_f \alpha \,,
      \end{equation}
   which tells us that it is unphysical. 

   Of course, the above is fully matched by the dynamical picture. 
   Dynamically, the vacuum 
    is achieved by the minimization of the 't Hooft determinant 
   (\ref{tDET}).
   In the minimum, the $\theta$-term is exactly cancelled by 
   the VEV of $\eta'$.
   Correspondingly, the generation of the $\eta'$ mass in 't Hooft language 
   is through the presence of the 't Hooft determinant and its non-zero VEV.  

     An alternative way of understanding the generation of 
     the $\eta'$ mass from TSV is via the Witten-Veneziano mechanism 
     \cite{Witten:1979vv, Veneziano:1979ec}.  Both languages can be  described as a $3$-form Higgs effect, in which the $\eta'$ is eaten up by the $3$-form and forms a massive pseudo-scalar
     \cite{Dvali:2005an}.

\subsubsection{Removing TSV via a gauge \texorpdfstring{$2$}{LG}-form}

 Let us now consider the case in which either there exists no anomalous  $U(1)$-symmetry or its quality is poor. 
 In such a case, the TSV can be removed by 
 the mechanism of $2$-form gauge axion
 introduced in  \cite{Dvali:2005an}.  
  
   The key ingredient is a $2$-form field  $B_{\mu\nu}$ that transforms under the $SU(N)$ gauge symmetry 
   (\ref{Cgauge}) as      
     \begin{equation} \label{Bgauge}
B_{\mu\nu} \, \rightarrow \, B_{\mu\nu} \, + \, \frac{1}{f}\Omega_{\mu\nu} \,,
\end{equation}
where $f$ is a scale.  
Notice that in this formulation $B_{\mu\nu}$ is positioned as 
an intrinsic part of the gauge redundancy, without 
any reference to a global symmetry. 

    Due to this, it enters the Lagrangian through the 
    following  gauge invariant combination,
   $\bar{C}^{\rm CS}_{\mu\nu\beta} \equiv C^{\rm CS}_{\mu\nu\beta} \, + \, f \, \partial_{[\mu}B_{\nu\beta]}$.  
   The lowest order term in the Lagrangian is: 
     \begin{equation} \label{CBgauge}
 \frac{1}{f^2}(C^{\rm CS}_{\mu\nu\beta} + f\partial_{[\mu}B_{\nu\beta]})^2 \,.
\end{equation}
 From this form, it is clear that the scale $f$ plays the 
 role of the cutoff.  Without loss of generality,  we can 
 tie it to the Planck mass  $f =M_P$. This case has the advantage of being protected by the gauge symmetry against arbitrary deformation to all orders in the operator expansion \cite{Dvali:2005an, Dvali:2023llt, Dvali:2022fdv, Sakhelashvili:2021eid}.    
  Correspondingly, in this formulation, the axion has exact quality.

   The vanishing of  TSV to all orders can be understood as the result of the $3$-form Higgs effect \cite{Dvali:2005an}. 
        Indeed, without $B_{\mu\nu}$, the TSV of the 
    $SU(N)$ is non-zero. 
   Therefore, $C^{\rm CS}$ contains a massless 
    $3$-form, $C$, 
   \begin{equation} \label{MofC}  
   C^{\rm CS} \, = \, \Lambda^2 \, C \, + \, {\rm heavy~ modes} \,,
  \end{equation} 
  where $\Lambda$ is the scale of TSV.
      All the $SU(N)$ fields can be integrated 
    out, and the resulting  EFT is,     
      \begin{equation} \label{CBEFT}
{\mathcal K}(E) \, + \,  \frac{1}{f^2}(\Lambda^2 C_{\mu\nu\beta} + f\partial_{[\mu}B_{\nu\beta]})^2 \,,
\end{equation}
 where ${\mathcal K}(E)$ is an algebraic 
 function of the fields strength $E \equiv \partial_{\alpha} 
C_{\mu\nu\beta} \epsilon^{\alpha\mu\nu\beta}$ \cite{Dvali:2005an} 
 \footnote{Notice that, since $E$ is an exterior derivative, 
it is the same for $\bar{C}^{\rm CS}_{\mu\nu\beta}$ and 
$C^{\rm CS}_{\mu\nu\beta}$.}. 
The higher-order derivative terms play no  
role in the vacuum structure and can be safely ignored. 
 Correspondingly, the vacuum derived from 
 (\ref{CBEFT}) is exact.

   It is clear that the $2$-form $B_{\mu\nu}$ acts as a St\"uckelberg  field for $C$, and the two combine into a massive
  $3$-form,  $\bar{C}^{\rm CS}_{\mu\nu\beta}$,
 which is equivalent to a massive 
 pseudo-scalar.  Obviously, this form is gauge invariant. 
 In the corresponding gauge invariant counterpart 
 of (\ref{CCcorr}) the would-be massless pole is shifted to 
 $p^2 =  \Lambda^4/f^2$ and TSV vanishes. 

 As already discussed previously, this construction 
 also shows that the existence of a massless pole in the 
 correlator (\ref{CCcorr}) of the original 
 theory with  non-zero TSV is a gauge-independent statement. 
 This massless pole can be recovered in the above gauge invariant 
 correlator of $\bar{C}^{\rm CS}_{\mu\nu\beta}$ by taking 
 the limit $f \rightarrow \infty$ in which the 
 St\"uckelberg $B_{\mu\nu}$ decouples.

\subsection{Electroweak sector} 
      
   We shall now apply the above understanding  to the 
   weak sector of the standard model.
   The main difference is that, unlike the color group of QCD, the 
    $SU(2)$ weak symmetry is in the Higgs phase, 
    due to a non-zero VEV of the 
   Higgs doublet $\langle \Phi \rangle =v$. 
 Despite this,  in the absence of fermions, 
 the TSV of  $SU(2)$ is non-zero.  Correspondingly, 
 the weak $\theta$ is physical \cite{Anselm:1992yz,Anselm:1993uj}. 
 In other words, the mass gaps of the  $SU(2)$ gauge bosons 
 generated by the Higgs, do not eliminate 
 the topological structure of the vacuum. Rather, they only constrain 
 the range of the instantons that effectively contribute to TSV
 (\ref{EEcorr}).  The effect is purely quantitative. 
 
 Correspondingly, as in QCD, 
 the $\theta$-vacuum structure
  of the $SU(2)$ weak sector is controlled by 
  TSV (\ref{EEcorr}) and correspondingly  
 by the pole at $p^2 =0$ in the Chern-Simons 
correlator  (\ref{CCcorr}).

   Let us now discuss the effect of fermions. 
 Per generation, they consists of four $SU(2)$-doublets of 
left-handed Weyl fermions:       
    three colors of the quark doublets and a single lepton doublet, 
 \begin{eqnarray}
q_L=\left(
\begin{array}{c}
u_L \\ 
d_L
\end{array} 
\right)~,~~
\ell_L=\left(
\begin{array}{c}
\nu_L \\ 
e_L
\end{array} 
\right)~. 
\label{doublets}
\end{eqnarray}
 In addition, the theory contains their right-handed counterparts 
 $u_R,~d_R,~e_R, \nu_R$, which are singlets of $SU(2)$. 
 The color and generation indexes are not shown explicitly. 
 
  We assume that all these fermions (including neutrinos) 
  form massive Dirac fermions due to their Yukawa couplings with the Higgs doublet $\Phi =\left(\phi^{+}, \phi^0\right)^{\rm T}$,   
  \begin{eqnarray} \label{YH}
\mathcal{L}=\nonumber\\
y_u\Phi \bar{q}_L u_R \,  + \,  y_{\nu}\Phi  \bar{\ell_L} \nu_R \,+ \, 
y_d\Phi^{c} \bar{q}_L d_R + y_e \Phi^{c}  \bar{\ell_L} e_R \,,
\end{eqnarray}
where $\Phi^{c}$ is a conjugated doublet. 
   
    Despite generating a mass gap, 
  these Yukawa couplings preserve the chiral 
  $U(1)_{B+L}$-symmetry, 
    \begin{eqnarray} \label{B+L}
   && (q_L, u_R, d_R)\, \rightarrow \, e^{i \alpha} \,(q_L, u_R, d_R) \,,  \nonumber \\  
    && (\ell_L, e_R, \nu_R)  \,\rightarrow e^{i 3\alpha} \, 
    (\ell_L, e_R, \nu_R) \,. 
\end{eqnarray}
 This symmetry is anomalous 
  with respect to $SU(2)$, resulting into a corresponding 
  shift of $\theta$. 
     
   Although $U(1)_{B+L}$-symmetry  has a good-quality 
   at the level of the Standard Model, we do not know whether 
   this quality holds in general.  In particular, it is not excluded 
   that  $U(1)_{B+L}$-symmetry is broken by high-dimensional 
   fermion operators. They generically appear in various  
extensions of the Standard Model. A well-known example 
of this sort is grand unification.  
    Since the nature of $\eta_{\rm w}$ depends on 
  the quality of $U(1)_{B+L}$, we consider the two options 
  separately.   
 
  \section{Good-quality  \texorpdfstring{$U(1)_{B+L}$}{LG}} 
  
   We start with the option that the quality 
   of $U(1)_{B+L}$ is good,  implying that the sole source 
   of its explicit breaking is the electroweak anomaly. In such a case $U(1)_{B+L}$  renders the electroweak $\theta$ unphysical \cite{Anselm:1992yz, Anselm:1993uj}\footnote{Notice that 
   an explicit breaking of either $B$ or $L$ symmetries (e.g., through the Majorana neutrino masses) does not change the situation qualitatively. The weak-$\theta$ remains unobservable due to the remaining $L$ or $B$ rotations, respectively.}. The effect is very similar to the one  
of a chiral symmetry of a massless quark on the $\theta$-term of QCD. In both cases, as a result of the anomalous symmetry, the TSV vanishes.

 Correspondingly, our general arguments apply to $U(1)_{B+L}$.  The elimination of weak-$\theta$ means that the massless pole in the correlator (\ref{CCcorr}) is removed.  
     As we already discussed, this implies the existence 
   of a new pseudo-scalar degree of freedom, $\eta_{\rm w}$,
   which realizes the $U(1)_{B+L}$-symmetry non-linearly.    
  The remaining question is the origin of this pseudo-Goldstone. 
   
 \subsection{External \texorpdfstring{$\eta_{\rm w}$}{LG}}  
       
 One option is that the $\eta_{\rm w}$-boson is an external degree 
 of freedom.  This possibility looks especially reasonable 
 in light of the fact that coupling to gravity
 was a crucial factor in justifying its existence. 
 The realization of this scenario is straightforward. 
 The $\eta_{\rm w}$-meson can be introduced as 
 a Goldstone phase degree of freedom of 
 a complex scalar field   $\Phi = |\Phi| e^{i\frac{\eta_{\rm w}}{f}}$
 that breaks $U(1)_{B+L}$-symmetry 
 spontaneously.  This is a full analog of  the 
 Peccei-Quinn scenario in QCD.  Such  generalizations of the 
 Peccei-Quinn axion has been considered previously in the 
 literature \cite{Anselm:1990uy, Dvali:2005an}.  
 
The field $\Phi$ can couple to quarks and leptons 
   via an arbitrary operator with non-zero 
   $U(1)_{B+L}$-charge and can be assigned an opposite 
   charge.   For example, the operator can be chosen 
   as, $\Phi \, q\, q\, q\, l$. In this case, under the $B+L$ 
   symmetry (\ref{B+L}) $\Phi$ transforms as 
   $\Phi \rightarrow e^{-i6\alpha} \Phi$. In other words, in this formulation good-quality 
   $U(1)_{B+L}$ is not a symmetry only of the Standard Model
   species but is necessarily shared with an external field 
   $\Phi$. 
   
\subsection{The fermion condensate} \label{cond_ch}

 The question that we now would like to ask is whether 
 the $\eta_{\rm w}$-boson could emerge from the electroweak 
 physics without any need for its extensions. 
  Of course, in such a case it can only emerge as a collective degree of freedom. 
    The  composite operator that matches the desired quantum numbers and the transformation properties is the 
phase of the 't Hooft determinant, which is generated by the 
$SU(2)_w$-instantons.  If this determinant would have a non-zero VEV,  the corresponding phase would be an obvious candidate for $\eta_{\rm w}$.   

 In order to gauge the plausibility of such a scenario, 
 we shall explicitly calculate the $(B+L)-$violating fermion condensate.  However,  
 we shall perform the calculation within a toy model 
that represents a simplified version of the electroweak sector
of the Standard Model. Namely, we shall get rid of the 
color and the hypercharge, thereby reducing the gauge sector 
to a weak $SU(2)_w$ group. Correspondingly, we reduce the fermion content to two
$SU(2)_w$-doublets of left-handed Weyl fermions  (\ref{doublets})
and their singlet right-handed partners, removing the color and generation quantum numbers. Basically, we shrink the fermion 
content of the Standard Model to a single generation of leptons and a single color of quarks. 
 The Yukawa sector of the Lagrangian is still described 
 by the (\ref{YH}), modulo the above simplification. 
 
Next, we switch to the Euclidean formulation. For convenience,  we combine fermions into an 8-component spinor $\Psi=\left(\psi, 
\phi\right)^{\rm T}$ \cite{Anselm:1992yz,Anselm:1993uj}, where
\begin{eqnarray}
\psi =q_L + \ell^c_R~,~~
\phi= 
\left(
\begin{array}{c}
u_R \\
d_R \\
\end{array} 
\right)+ 
\left(
\begin{array}{c}
e^c_L \\ 
-\nu^c_L
\end{array} 
\right)~. 
\label{comb}
\end{eqnarray}
Note, $\psi$ comprises of gauge $SU(2)$-doublet spinors, while $\phi$ contains $SU(2)$-singlet ones. 

With the above notations, the fermionic Lagrangian in Euclidean space can be written in a compact form as\footnote{We recall that when turning to the Euclidean space, a  Minkowski spinor $\bar\phi \to -i\phi^{\dagger}$ ($\bar\phi_{L,R} \to -i\phi^{\dagger}_{R,L}$), whereas the Euclidean spinor $\phi^{\dagger}$ 
is an independent field rather than a complex
conjugated field to $\phi$. The Euclidean gamma matrices are defined such that $\lbrace \gamma_{\mu}, \gamma_{\nu}\rbrace=2\delta_{\mu\nu}$. For further conventions, see, e.g., \cite{Shifman:2022shi}. }:
\begin{equation}
{\cal L}_F=\Psi^{\dagger} \hat {\cal D} \Psi~,
\label{lagrang}
\end{equation}
where $\hat {\cal D}$ is given by, 
\begin{eqnarray}
\left(
\begin{array}{cc}
-i\slashed{D}& i\epsilon M^{*}_{\ell}\epsilon P_L-iM_qP_R\\
i\epsilon M^{\rm T}_{\ell}\epsilon P_R-iM_q^{\dagger}P_L & -i\slashed{\partial}
\end{array} 
\right).
\label{oper}
\end{eqnarray}
 Here, $\slashed{D}=\gamma_{\mu}\left(\partial_{\mu}-i\hat W_{\mu}\right)$; $P_{L.R}=(1 \pm \gamma_5)/2$  and $M_q$ and $M_{\ell}$ embody the quark-Higgs and lepton-Higgs Yukawa interactions, respectively:
\begin{eqnarray}
M_q=\left(
\begin{array}{cc}
y_u\phi^{0*}~,& y_d\phi^+ \\
-y_u\phi^{+*}~, & y_d\phi^0
\end{array} 
\right);
\label{Yukawa}
\end{eqnarray}
$M_{\ell}$ is obtained from $M_q$ by replacing the quark Yukawa couplings with the corresponding leptonic ones: $y_{u,d}\to y_{\nu, e}$. 

The  $U(1)_{B+L}$-symmetry (\ref{B+L})
of the full Standard Model, with a proper normalization,  translates 
as the following global transformation:
\begin{equation}
\Psi \to {\rm e}^{i\alpha \Gamma_5/2}\Psi~,~~\Psi^{\dagger}\to \Psi^{\dagger}{\rm e}^{i\alpha \Gamma_5/2}~,
\label{transform}
\end{equation}
which leaves the 
Lagrangian (\ref{lagrang}) invariant.
The generalised chirality operator has the form:
\begin{eqnarray}
\Gamma_5=\left(
\begin{array}{cc}
\gamma_5~,& 0 \\
0~, & -\gamma_5
\end{array}
\right)\,.
\label{chiral}
\end{eqnarray}
Hence, both left-handed and right-handed quarks and leptons have positive $\Gamma_5$ chirality, while their anti-particles carry the negative $\Gamma_5$ chirality. 

The non-perturbative sector in our model, like in the full electroweak theory, is dominated by $\nu=\pm 1$ one (anti)instanton contributions. These are exponentially suppressed relative to the topologically trivial sector, which is dominated by the perturbative physics. This is because in the Higgs phase, the weak-$SU(2)$ instantons are constrained and thereby are screened at distances larger than
the electroweak length $\sim 1/v$ \cite{Affleck:1980mp,tHooft:1976snw}. 

As long as the theory remains weakly coupled, the non-interacting ideal instanton gas is expected to be an excellent approximation. The explicit field configurations that describe the constrained (anti)instantons with the unit topological charge can be found in \cite{Shifman:2022shi}. Also, notice that for non-zero Yukawa couplings, $y_{u,d}, y_{\nu, e}$, the theory is fully gapped, with no massless degrees of freedom. However, the massive fermions nevertheless exhibit normalizable zero modes in the background of the electroweak instantons \cite{Krasnikov:1979kz}.

Within this setup, the $B+L-$violating fermionic condensate $\langle \Psi^{\dagger}(x)\Psi(x)\rangle$ can be straightforwardly computed. It is represented by the fermion propagator in the background of the instanton gas, $\left(\hat{\cal D}+i\mu\right)^{-1}$ averaged over the instanton configurations and positions as well as zero and massive modes of gauge, Higgs and fermion fields\footnote{In dealing with integration over fermions we introduce a regulator parameter $\mu$, by adding $(B+L)-$violating term, $i\mu \Psi^{\dagger}\Psi$, to the Lagrangian. It must be taken to $0$ at the final stage of computation.}: 
\begin{eqnarray}
\langle \Psi^{\dagger}(x)\Psi(x)\rangle =\nonumber\\=\lim_{\mu\to 0}{\int \frac{d^4zd\rho}{\rho^5}}~D(\rho)\langle x\vert \left(\hat{\cal D}+i\mu\right)^{-1} \vert x \rangle.
\label{vev}
\end{eqnarray}
In the above equation, the quantity 
\begin{eqnarray}
D(\rho)=\left(\frac{2\pi}{\alpha(\rho)}\right)^4{\rm e}^{-\frac{2\pi}{\alpha(\rho)}-2\pi^2v^2\rho^2}~\rho\mu ~,
\end{eqnarray}
is interpreted as the density of instantons of size $\rho$ in the presence of fermion zero modes and $\alpha(\rho)$ is the effective $SU(2)$ gauge coupling constant evaluated at the scale $\rho$. As discussed, the weak-$\theta$ is absorbed in the condensate phase and omitted here. 

In order to show that the above condensate is indeed non-zero, we inspect the fermion propagator $\left(\hat {\cal D}+i\mu\right)^{-1}$ and separate it into the propagator for zero modes $P_0$ and the propagator for massive modes $\Delta$ \cite{Brown:1977eb}:    
\begin{equation}
\frac{1}{\hat{\cal D}+i\mu } = \frac{P_0}{i\mu} + \Delta - i\mu \Delta^2 +{\cal O}(\mu^2)~. 
    \label{prop}
\end{equation} 
Plugging this propagator into Eq. (\ref{vev}), we observe that the only contribution that survives the $\mu \to 0$ limit comes from the fermion zero modes. After evaluating the integrals, we obtain:      
\begin{eqnarray}
\langle \Psi^{\dagger}(x)\Psi(x)\rangle &\simeq&-iv^3\left(\frac{2\pi}{\alpha}\right)^4{\rm e}^{-\frac{2\pi}{\alpha}}\,.
\label{vev1}
\end{eqnarray}
In the course of the above calculations, we have used $\langle x\vert P_0\vert x \rangle = \Psi_0^{\dagger}(x-z)\Psi_0(x-z)$ and the normalisation of the zero mode wavefunction: $\int d^4x \Psi_0^{\dagger}(x)\Psi_0(x)=1$. The condensate in Minkowski space is related to the one calculated in Euclidean space as $\langle \Bar\Psi\Psi\rangle=-i\langle \Psi^{\dagger}\Psi\rangle$ and hence is real. 

The fermion condensate 
provides an internal mechanism for 
spontaneous breaking of $U(1)_{B+L}$-symmetry. 
The corresponding pseudo-Goldstone excitation, $\eta_{\rm w}$, 
is the right candidate for eliminating the massless pole in the 
correlator (\ref{CCcorr}). In order to see this explicitly, let us consider the following anomalous Ward identity (see, e.g.,  \cite{Shore:2007yn} in the context of QCD) that follows from evaluating the vacuum expectation value of the variation $\delta(\Psi^+\Gamma_5\Psi)=2i\Psi^+\Psi$ under the generalised chiral transformations (\ref{transform}):
\begin{eqnarray}
&\int d^4x
\left\langle \left(i\mu \Psi^+\Gamma_5\Psi - \frac{\alpha}{4\pi}F\tilde F\right) (x)~,~\Psi^+\Gamma_5\Psi (0) \right\rangle =& \nonumber \\
&i\langle \Psi^+\Psi\rangle \neq 0.&
\label{ward}
\end{eqnarray}
The first term on the lhs of (\ref{ward}) comes from the variation of the classical Lagrangian, while the second term originates from the anomaly. In the absence of this anomalous term, using the relation  $\partial_{\mu}J_{B+L}^{\mu}=2i\mu \Psi^+\Gamma_5\Psi$,  one immediately infers the existence of the massless pole corresponding to the Goldstone boson. The anomaly contribution ensures that no massless particles exist in the spectrum. Taking $\mu\to 0$ in (\ref{ward}) we obtain:
\begin{eqnarray}
&\int d^4x\langle  F\tilde F(x)~,~ \Psi^+\Gamma_5\Psi \rangle_{p=0} \propto \langle \Psi^+\Psi\rangle.
\label{ward2}
\end{eqnarray}
It is therefore clear that for satisfying the above identity, there must exist a state $\vert \eta\rangle$ such that $\langle 0\vert F\tilde F\vert \eta\rangle=B(p)\neq 0$ and $\langle \eta\vert \Psi^+ \Gamma_5\Psi\vert 0\rangle=C(p)\neq 0$. In turn, this implies that the topological susceptibility contains a massive pole in its K\"all\'en-Lehmann spectral decomposition. Equivalently, the Chern-Simons 3-form 
becomes a massive propagating field \cite{Dvali:2005an}:
\begin{equation}
FT\langle C^{(\rm{CS})},C^{(\rm{CS})}\rangle =  \frac{\rho(0)}{p^2-m_\eta^2} +..., 
    \label{Cmass}
\end{equation}
where $\rho(0)=\vert B(0)\vert^2\neq 0$.

A comment on the validity of the dilute instanton gas approximation employed in the above calculations is in order. It is certainly valid at energy scales below the sphaleron threshold $\sim 2\pi M_W/\alpha$. Above this threshold, multiple vector boson exchanges between instantons must be takewn into account, and the series must be resummed. In this regime, the process likely becomes dominated by the instanton--anti-instanton bound states rather than the individual non-interacting instantons. Since such bound states carry the trivial topological charge, we expect fermion zero modes to delocalize and hence the fermion condensate to ``evaporate". We thus may regard the sphaleron threshold
as an absolute upper bound on the ultraviolet scale beyond which the emergent $\eta_{\rm w}$ does not exist and the relevant effects are described by multi-particle states.

\subsection{Physical meaning and validity of the fermion condensate}

 Let us reflect on the meaning of the above computation 
for our proposal.   First,  we can take it as an indication 
that a gauge theory that is in the Higgs phase, in principle,  
could accommodate the required $\eta_{\rm w}$-type particle.  
Namely, if the condensate exists, 
$\eta_{\rm w}$ emerges as its phase. As already noted, this is 
somewhat analogous to the emergence of the 
$\eta'$-meson as of the phase of the quark condensate 
in ordinary QCD.  

However, this analogy must be
weighted very carefully.  
For one,  in the present case,
the theory is in the Higgs phase, and fermions are not confined. This creates a set of questions. In particular, if $\eta_{\rm w}$ is a collective mode, the domain of validity of its EFT must impose further restrictions on the parameters, such as the Yukawa couplings. What these conditions are and whether they
can be satisfied within a more realistic setup, has to be studied 
separately. 

 Another obvious question is what is the role of gravity? On one hand,  in the case of an good-quality $U(1)_{B+L}$-symmetry, 
 gravity demands the existence $\eta_{\rm w}$ in form of a 
 pseudo-Goldstone boson of  $U(1)_{B+L}$. 
 However, if the condensate can be provided 
 entirely by the non-perturbative electroweak $SU(2)$ dynamics,  the emergence of 
$\eta_{\rm w}$  appears to be guaranteed without gravity. Such a scenario exhibits no {\it a priori} inconsistency. Indeed, if an interacting $\eta_{\rm w}$ exists already in the limit of $M_P = \infty$, it easily accommodates the demands 
of gravity also for a finite $M_P$. However,  at least at the level of EFT,  such a scenario appears to be a lucky coincidence in which the consistency demands of gravity are met by the EFT already for $M_P =\infty$.  

Notably, similar precedents do exist.  An example is 
the cancellation of the gravitational anomalies within the low-energy EFT.   Such cancellations must take place for arbitrary values of $M_P$, including $M_P =\infty$. 

A more specific example, which is directly relevant to the present case, is the demand for the cancellation of the chiral gravitational anomaly among the spin-$1/2$ fermions  \cite{Dvali:2024dlb}\footnote{On a separate note, this requirement can have interesting phenomenological implications for Standard Model neutrinos, such as  
the presence of their right-handed partners.}. As argued in the latter work, this condition is 
imposed by the Eguchi-Hanson instantons \cite{Eguchi:1978xp, Eguchi:1978gw} since such fermions give no zero modes in their background \cite{index, Atiyah:1975jf}. Due to this, the chiral gravitational anomaly must be taken up by a spin-$3/2$ fermion. This leads to the formation of their condensate \cite{Hawking:1978ghb, Konishi:1988mb}, which breaks the anomalous $R$-symmetry spontaneously.   The $R$-axion emerges as the phase of the gravitino condensate. 
  
 Despite some striking similarities,  the above precedent cannot be directly transported 
  to the case of the  Higgsed gauge symmetries and especially 
 to the Standard Model.  Even if in a toy model, in which the fermion masses are free parameters, the fermion condensation 
 takes place, it is far from being clear whether this is 
 possible within the Standard Model. \footnote{Further analysis performed since the submission of this paper indicates  the validity of the fermion condensate in the full  Standard Model.}  
 
Once again, we would like to stress that the question of the possible emergence of $\eta_{\rm w}$ from the fermion condensate changes nothing about the necessity of its existence and its finite-strength coupling to the electroweak TSV at finite $M_P$.

 \section{Poor quality \texorpdfstring{$B+L$}{LG} and  \texorpdfstring{$2$}{LG}-form 
    \texorpdfstring{$\eta_{\rm w}$}{LG}} 
   
   So far, we have been considering the situation in which 
   the explicit breaking of the $U(1)_{B+L}$-symmetry was coming  
   exclusively from the $SU(2)$-instantons. 
   In such a case, irrespective of its precise origin, the particle $\eta_{\rm w}$ represents a pseudo-Goldstone boson of $B+L$-symmetry. 
   
   {  We shall now discuss how the situation changes if the $U(1)_{B+L}$-symmetry is 
   of poor quality.  That is, if we assume that this symmetry is explicitly broken 
   beyond the electroweak anomaly. We shall split the discussion in the following two stages. 
   
   First, without 
   any extra assumptions, we show that the existence of 
   $\eta_{\rm w}$ is not jeopardized by the quality of $U(1)_{B+L}$-symmetry. Next, we show that by employing certain
    well-justified assumptions about gravity, one can 
    go further and motivate a gauge formulation of $\eta_{\rm w}$. } \\  

{\it   Insensitivity of $\eta_{\rm w}$ towards $B+L$-breaking} \\
   
  {  In order to understand the effect of this 
  breaking on  $\eta_{\rm w}$, we notice that the explicit breaking of $U(1)_{B+L}$ requires physics from BSM, since, as it is well-known, among the Standard Model fields there exist no gauge invariant renormalizable operators that can break $U(1)_{B+L}$. At the same time, we know from the existing experimental constraints that
  $B+L$-violating high-dimensional operators must be extremely suppressed.  Equipped with this knowledge, it is easy to see that the existence of $\eta_{\rm w}$-meson, derived in the limit
of exact quality $B+L$-symmetry, is insensitive towards
any phenomenologically-acceptable explicit breaking of this symmetry.

  For example, the $U(1)_{B+L}$-symmetry can be explicitly broken by dimension-$6$ operators of the type, 
   \begin{equation} \label{qqql}
     \frac{1}{M^2} qqql \,,   
   \end{equation}
where $M$ is the scale of some BSM physics. 
 The well-known phenomenological constraints on baryon and lepton number violating processes imply that the scale 
$M$ has to be extremely high,  $M > 10^{16}$GeV. This immediately tells us that the effect of such operators 
 on $\eta_{\rm w}$-meson is totally negligible. For example, 
 the relative correction to the fermion condensate
 calculated for $M = \infty$ (i.e., for exact quality $B+L$)
 is suppressed by powers of the scale  $M$. 
 It is obvious that, for all practical purposes, this correction is irrelevant.  

 We thus conclude that the explicit breaking of  $U(1)_{B+L}$-symmetry cannot jeopardize the very existence of the 
 $\eta_{\rm w}$-meson deduced in  the limit of exact quality 
 $B+L$-symmetry. } \\

 {\it   The case from gravity for gauge realization of $\eta_{\rm w}$} \\

{ There exists an additional (and fully independent) reasoning which in the 
case of a poor quality $B+L$-symmetry not only suggests the necessity 
of the $\eta_{\rm w}$-meson but also makes the case for its gauge realization. This reasoning is based on more 
fundamental knowledge about gravity. Although it relies on different assumptions, it does not affect the previous 
conclusion about the existence of $\eta_{\rm w}$. 
Rather, it provides additional supporting 
evidence for it and at the same 
time makes the case for its fully gauge invariant realization. This reasoning goes as follows. 

 In the case of a poor-quality $B+L$-symmetry,  the TSV 
 cannot be fully nullified by the anomalous symmetry
 which is explicitly broken beyond the anomaly 
 by the operators of the type (\ref{qqql}). 
 Correspondingly, the $\theta$-vacua, although highly suppressed in energy, remain physical. } 
  
This, however, is incompatible with the 
   $S$-matrix formulation of gravity \cite{Dvali:2018dce, Dvali:2022fdv, Dvali:2023llt}.
  This is due to the fact  \cite{Dvali:2020etd}
  that the $S$-matrix formulation, currently the only existing formulation of quantum gravity, is incompatible 
  with the vacua with non-asymptotically flat cosmologies.
   These include the de Sitter vacua \cite{Dvali:2013eja, Dvali:2014gua, Dvali:2017eba} as well as any anti-de Sitter type 
   vacua leading to big crunch cosmologies
   \cite{Dvali:2022fdv, Dvali:2023llt}. Correspondingly, the $\theta$-vacua must be eliminated. 
   However, this is not possible via a $U(1)_{B+L}$-Goldstone, since this symmetry is of poor quality.   
 Instead, the goal has to be achieved via the introduction of $\eta_{\rm w}$ in the form of a gauge axion
   \cite{Dvali:2005an}.   
  
  As we already discussed, in this scenario, we introduce $\eta_{\rm w}$ as a $2$-form field  $B_{\mu\nu}$ that transforms under the 
  $SU(2)$ gauge symmetry  (\ref{Cgauge})  as (\ref{Bgauge}). 
 The construction follows the steps that were already
 outlined for the generic $SU(N)$-symmetry. The $2$-form $B_{\mu\nu}$ plays the role of the St\"uckelberg field 
  that compensates the $SU(2)$-gauge shift (\ref{Cgauge}) of  
  $C$.  This puts $C$ in the Higgs phase, shifting the 
  pole in the correlator (\ref{CCcorr}) to $p^2 \neq 0$. Due to the  protection by the gauge symmetry, (\ref{Cgauge}),  (\ref{Bgauge}), the $B_{\mu\nu}$ realization of $\eta_{\rm w}$ guarantees that the electroweak $\theta$-vacuum is nullified to all orders in operator expansion \cite{Dvali:2005an, Sakhelashvili:2021eid,  Dvali:2022fdv, Dvali:2023llt}.  

  Notice that although at the level of the low-energy EFT, the
  $B_{\mu\nu}$-formulation can be dualized into a pseudo-scalar axion
  with  an arbitrary potential \cite{Dvali:2005an}, the 
  UV-completion of the latter in the form of a Goldstone phase of a complex field  breaks duality, explaining the stability of $B_{\mu\nu}$-quality relative to the Goldstone case \cite{Dvali:2022fdv}.

\section{The double role of gravity}
 
   We are learning that gravity provides a dual motivation 
   for the existence of $\eta_{\rm w}$.   The concrete arguments 
   depend on the status of $U(1)_{B+L}$-symmetry and on how far one is willing to take the gravitational constraints.

{ At most conservative level of reasoning, it suffices to employ the very basic feature of gravity: its universal nature. This ensures that at a finite value of 
 $M_P$ no fully decoupled fields can exist in the theory.
 As already said, since this feature is a direct consequence of general covariance, we regard it as a common sense  rather 
 than an extra assumption. 
 
 However, for the full clarity, let us elaborate on this topic.  It is well-known that by the power of general covariance, the Einstein gravity couples to all canonically normalized fields with a universal strength set by $1/M_P$. In particular, this applies to a massless 
 $3$-form field. Gravitational effects of this field
 are well-understood (see, e.g.,
 \cite{Brown:1988kg, Dvali:2003br, Dvali:2004tma, Dvali:2005zk}). Applied to the present case, this feature guarantees the impossibility of decoupling of a massless $p^2=0$ pole from the  TSV of a gauge theory. 
 
Notice that this non-decoupling argument is very different from the so-called 
weak gravity conjecture \cite{Arkani-Hamed:2006emk}, which independently leads 
to the same conclusion.  Indeed, this conjecture implies that 
gravity must be the weakest force.  Although, initially formulated for abelian $U(1)$ gauge forces, if applied to the 
force mediated by a massless $3$-form, it would also 
exclude the possibility of decoupling at finite 
$M_P$.

 As already explained, the impossibility of decoupling 
 of the $3$-form,  leaves the only option 
 for vanishing of the  TSV: the shift of the pole towards 
 a non-zero mass, $p^2 = m_{\eta}^2$. } This leads to the existence of $\eta_{\rm w}$  as of 
 $U(1)_{B+L}$ pseudo-Goldstone.

  {  The existence of $\eta_{\rm w}$ is insensitive to the quality of 
  $B+L$ symmetry and persists even if the symmetry is of 
   poor quality, i.e., it is 
  explicitly broken by the sources beyond the $SU(2)$-anomaly.   For example, the explicit breaking can originate from higher dimensional operators of the type (\ref{qqql})
  
  As already discussed, due to its extremely suppressed strength, such breaking cannot jeopardize the existence of $\eta_{\rm w}$, since the corrections to the effects generated by the electroweak instantons are totally negligible.
  Correspondingly, the pseudo-Goldstone nature of $\eta_{\rm w}$ 
  derived for $M=\infty$ is unaffected. 

 What changes, however, is that the pseudo-Goldstone $\eta_{\rm w}$ cannot fully cancel the electroweak $\theta$-term. 
 In particular, this is clear from the fact that 
 the fermionic zero modes in the instanton background become gapped.  Correspondingly, the $\theta$-vacuum, although extremely suppressed, is physical.

 As explained, by employing the additional requirement of a well-defined $S$-matrix,  we can obtain further restrictions on the nature of $\eta_{\rm w}$.
 Since in this case the 
 the quality requirement is exact, it is most natural that 
  $\eta_{\rm w}$ is introduced as the $2$-form 
  $B_{\mu\nu}$, which transforms as a St\"uckelberg under the 
  $SU(2)$-gauge symmetry \cite{Dvali:2005an}.  In this case, no anomalous global symmetry is required. The $\theta$-vacua of $SU(2)$-theory are absent in all orders in the operator expansion. }

 A few comments are in order. First, depending on the 
 level of $U(1)_{B+L}$-quality, external $\eta_{\rm w}$ can 
 coexist with the phase of the condensate and mix with 
 it. This is similar to the mixing between the hidden axion 
 and the $\eta'$-meson in QCD. Secondly, gravity demands that the mass of the proper $\eta_{\rm w}$ must be generated exclusively from the electroweak TSV.  This follows from the 
   $S$-matrix requirement of exact vanishing of TSV \cite{Dvali:2022fdv}, as well as, from the requirement that 
   spin-$1/2$ fermions must not contribute to the gravitational anomaly due to the absence of their zero modes in the 
   Eguchi-Hanson instantons \cite{Dvali:2024dlb}.

\section{Discussions}

{
In conclusion, in the present paper, we have provided 
different evidences that lead us, as the only  
rational explanation, to the prediction of a new particle, $\eta_{\rm w}$-boson, in the electroweak sector of the standard model. The key ingredients are the presence of the massive pole in the TSV correlator on one hand and the presence 
of the B+L-violating condensate with matching quantum numbers on the other.  Both ingredients were achieved
by using the well-accepted tools such as, the Ward identities, 
gauge invariance and instanton calculus, without going 
beyond the validity of approximations used (e.g., such as the dilute instanton gas). At the same time, gravity played the role of an additional 
useful ``spectator" tool that enabled the monitoring of the non-decoupling of the pole in TSV. 

In the light of our analysis, 
under the above conservative assumptions that are organic for the currently accepted framework, the existence of $\eta_{\rm w}$-boson comes up as a matter of consistency for the topological structure of the Standard Model vacuum and its embedding in gravity. }

 A highly intriguing possibility is the 
   emergence of  $\eta_{\rm w}$ in the form of the phase of the 
  $U(1)_{B+L}$-violating  fermionic condensate of quarks and leptons. As a step in this direction, we performed an illustrative computation in a toy version of the Standard Model, which appears to support the generation of the fermion condensate by the instantons. 
   However, several open questions remain. Namely, the role of gravity, the range of validity of EFT of $\eta_{\rm w}$, as well as the extrapolation of the results to a fully realistic version of the Standard Model, must be further scrutinized.        

The phenomenological and cosmological implications of $\eta_{\rm w}$ depend on its decay constant $f$. Since the mass of $\eta_{\rm w}$ is generated from the electroweak TSV, for all the reasonable values of the  scale $f$, 
$\eta_{\rm w}$ is expected to be an extraordinarily light and weakly interacting particle. 
In the case of emergent $\eta_{\rm w}$, the phenomenological relevant parameters can, in principle, be extracted from 2-instanton correlators, employing the formalism of \cite{Pisarski:2019upw}. 

Irrespective of the actual origin, the (pseudo)Goldstone nature of $\eta_{\rm w}$ and the anomaly relations dictate its interactions with electroweak gauge bosons, which would be suppressed by $\propto 1/f$. Note, however, that coupling of $\eta_{\rm w}$ to two photons is absent due to the absence of the mixed $B+L$-electromagnetic anomaly. The interactions with fermions that are generated non-perturbatively are further suppressed by a large exponential factor stemming from instantons. Therefore, the most significant coupling of $\eta_{\rm w}$ to quarks and leptons emerges at (effectively) a two-loop level through the loop of massive electroweak gauge bosons. This may induce rare decays of mesons with the characteristic signature of missing energy carried away by $\eta_{\rm w}$. These processes can be distinguished from the Standard Model background processes where the missing energy is carried away by neutrinos if $f$, as we expect, is not far from the electroweak scale.

{ Finally, going beyond the minimal knowledge and making more 
advanced (although fully justified) assumptions, such as the necessity of a well-defined $S$-matrix in gravity, allows 
us to further restrict the nature of $\eta_{\rm w}$
motivating its gauge formulation  \'a la gauge-axion \cite{Dvali:2005an}. However, the level of minimalism in the above two steps must be distinguished very clearly. 
}

\section*{Acknowledgments}
We would like to thank Misha Shaposhnikov for discussions on fermionic zero modes and condensate, and Lasha Berezhiani for the discussion on the validity of the instanton gas approximation. OS would like to thank the Arnold Sommerfeld Center at LMU, the Max-Planck-Institute for Physics, Munich and MITP Mainz (Cluster of Excellence PRISMA+ Project ID 390831469) for the hospitality extended during the completion of this work. The work of GD was supported in part by the Humboldt Foundation under the Humboldt Professorship Award, by the European Research Council Gravities Horizon Grant AO number: 850 173-6, by the Deutsche Forschungsgemeinschaft (DFG, German Research Foundation) under Germany’s Excellence Strategy - EXC-2111 - 390814868, Germany’s Excellence Strategy under Excellence Cluster Origins 
EXC 2094 – 390783311. The work of AK and OS was partially supported by the Australian Research Council under the Discovery Projects grants DP210101636 and DP220101721.     \\

\noindent {\bf Disclaimer:} Funded by the European Union. Views and opinions expressed are, however, those of the authors only and do not necessarily reflect those of the European Union or European Research Council. Neither the European Union nor the granting authority can be held responsible for them.\\


\begin{thebibliography}{99} 
\bibitem{Dvali:2005an}
G.~Dvali,
``Three-form gauging of axion symmetries and gravity,''
[arXiv:hep-th/0507215 [hep-th]].


\bibitem{Dvali:2018dce}
G.~Dvali, C.~Gomez and S.~Zell,
``A Proof of the Axion?,''
[arXiv:1811.03079 [hep-th]].


\bibitem{Dvali:2022fdv}
G.~Dvali,
``Strong-$CP$ with and without gravity,''
[arXiv:2209.14219 [hep-ph]].


\bibitem{Dvali:2023llt}
G.~Dvali,
``The role of gravity in naturalness versus consistency: strong-CP and dark energy,''
Phil. Trans. Roy. Soc. Lond. A \textbf{382}, no.2266, 20230084 (2023).


\bibitem{Dvali:2005ws}
G.~Dvali, R.~Jackiw and S.~Y.~Pi,
``Topological mass generation in four dimensions,''
Phys. Rev. Lett. \textbf{96}, 081602 (2006)
[arXiv:hep-th/0511175 [hep-th]].



\bibitem{Dvali:2013cpa}
G.~Dvali, S.~Folkerts and A.~Franca,
``How neutrino protects the axion,''
Phys. Rev. D \textbf{89}, no.10, 105025 (2014)
[arXiv:1312.7273 [hep-th]].


\bibitem{Dvali:2016uhn}
G.~Dvali and L.~Funcke,
``Small neutrino masses from gravitational \ensuremath{\theta}-term,''
Phys. Rev. D \textbf{93}, no.11, 113002 (2016)
[arXiv:1602.03191 [hep-ph]].


\bibitem{Dvali:2017mpy}
G.~Dvali,
``Topological Origin of Chiral Symmetry Breaking in QCD and in Gravity,''
[arXiv:1705.06317 [hep-th]].


\bibitem{Hebecker:2019vyf}
A.~Hebecker and P.~Henkenjohann,
``Gauge and gravitational instantons: From 3-forms and fermions to Weak Gravity and flat axion potentials,''
JHEP \textbf{09}, 038 (2019)
[arXiv:1906.07728 [hep-th]];

and G.~Dvali, private communications therein.   


\bibitem{Peccei:1977hh}
R.~D.~Peccei and H.~R.~Quinn,
``CP Conservation in the Presence of Instantons,''
Phys. Rev. Lett. \textbf{38} (1977), 1440-1443


\bibitem{Weinberg:1977ma}
S.~Weinberg,
``A New Light Boson?,''
Phys. Rev. Lett. \textbf{40} (1978), 223-226

\bibitem{Wilczek:1977pj}
F.~Wilczek,
``Problem of Strong  $P$  and  $T$  Invariance in the Presence of Instantons,''
Phys. Rev. Lett. \textbf{40} (1978), 279-282

\bibitem{tHooft:1976snw}
G.~'t Hooft,
``Computation of the Quantum Effects Due to a Four-Dimensional Pseudoparticle,''
Phys. Rev. D \textbf{14}, 3432-3450 (1976)
[erratum: Phys. Rev. D \textbf{18}, 2199 (1978)].

\bibitem{Witten:1979vv}
E.~Witten,
``Current Algebra Theorems for the U(1) Goldstone Boson,''
Nucl. Phys. B \textbf{156}, 269-283 (1979).

\bibitem{Veneziano:1979ec}
G.~Veneziano,
``U(1) Without Instantons,''
Nucl. Phys. B \textbf{159}, 213-224 (1979).



\bibitem{Sakhelashvili:2021eid}
O.~Sakhelashvili,
``Consistency of the dual formulation of axion solutions to the strong CP problem,''
Phys. Rev. D \textbf{105}, no.8, 085020 (2022).
[arXiv:2110.03386 [hep-th]].

\bibitem{Anselm:1992yz}
A.~A.~Anselm and A.~A.~Johansen,
``Baryon nonconservation in standard model and Yukawa interaction,''
Nucl. Phys. B \textbf{407}, 313-330 (1993).

\bibitem{Anselm:1993uj}
A.~A.~Anselm and A.~A.~Johansen,
``Can electroweak theta term be observable?,''
Nucl. Phys. B \textbf{412}, 553-573 (1994)
[arXiv:hep-ph/9305271 [hep-ph]].


\bibitem{Anselm:1990uy}
A.~A.~Anselm,
``Periodic universe and condensate of pseudoGoldstone field,''
Phys. Lett. B \textbf{260}, 39-44 (1991).

\bibitem{Shifman:2022shi}
M.~Shifman,
``Advanced Topics in Quantum Field Theory,''
Cambridge University Press, 2022.
ISBN 978-1-108-88591-1, 978-1-108-84042-2


\bibitem{Affleck:1980mp}
I.~Affleck,
``On Constrained Instantons,''
Nucl. Phys. B \textbf{191}, 429 (1981)

\bibitem{Krasnikov:1979kz}
N.~V.~Krasnikov, V.~A.~Rubakov and V.~F.~Tokarev,
``Zero fermion modes in models with spontaneous symmetry breaking,''
J. Phys. A \textbf{12}, L343-L346 (1979).

\bibitem{Brown:1977eb}
L.~S.~Brown, R.~D.~Carlitz, D.~B.~Creamer and C.~k.~Lee,
``Propagation Functions in Pseudoparticle Fields,''
Phys. Rev. D \textbf{17}, 1583 (1978).

\bibitem{Shore:2007yn}
G.~M.~Shore,
``The U(1)(A) Anomaly and QCD Phenomenology,''
Lect. Notes Phys. \textbf{737}, 235-288 (2008)
[arXiv:hep-ph/0701171 [hep-ph]].

\bibitem{Dvali:2024dlb}
G.~Dvali, A.~Kobakhidze and O.~Sakhelashvili,
``Hint to Supersymmetry from GR Vacuum,''
[arXiv:2406.18402 [hep-th]].

\bibitem{Eguchi:1978xp} 
T.~Eguchi and A.~J.~Hanson,
``Asymptotically Flat Selfdual Solutions to Euclidean Gravity,''
Phys. Lett. B \textbf{74}, 249-251 (1978).


\bibitem{Eguchi:1978gw}
T.~Eguchi and A.~J.~Hanson,
``Selfdual Solutions to Euclidean Gravity,''
Annals Phys. \textbf{120}, 82 (1979).


\bibitem{index} 
 M.~F.~Atiyah and I.~M.~Singer, ``The index of elliptic operators on compact manifolds,'' Bull. Amer. Math.
Soc. {\bf 69}, 422–433 (1963). 
\bibitem{Atiyah:1975jf}
M.~F.~Atiyah, V.~K.~Patodi and I.~M.~Singer,
``Spectral asymmetry and Riemannian Geometry 1,''
Math. Proc. Cambridge Phil. Soc. \textbf{77}, 43 (1975).

\bibitem{Hawking:1978ghb}
S.~W.~Hawking and C.~N.~Pope,
``Symmetry Breaking by Instantons in Supergravity,''
Nucl. Phys. B \textbf{146}, 381-392 (1978).

\bibitem{Konishi:1988mb}
K.~Konishi, N.~Magnoli and H.~Panagopoulos,
``Spontaneous Breaking of Local Supersymmetry by Gravitational Instantons,''
Nucl. Phys. B \textbf{309}, 201 (1988).

\bibitem{Dvali:2020etd}
G.~Dvali,
``$S$-Matrix and Anomaly of de Sitter,''
Symmetry \textbf{13}, no.1, 3 (2020)
[arXiv:2012.02133 [hep-th]].

\bibitem{Dvali:2013eja}
G.~Dvali and C.~Gomez,
``Quantum Compositeness of Gravity: Black Holes, AdS and Inflation,''
JCAP \textbf{01}, 023 (2014)
[arXiv:1312.4795 [hep-th]].


\bibitem{Dvali:2014gua}
G.~Dvali and C.~Gomez,
``Quantum Exclusion of Positive Cosmological Constant?,''
Annalen Phys. \textbf{528}, 68-73 (2016)
[arXiv:1412.8077 [hep-th]].


\bibitem{Dvali:2017eba}
G.~Dvali, C.~Gomez and S.~Zell,
``Quantum Break-Time of de Sitter,''
JCAP \textbf{06}, 028 (2017)
[arXiv:1701.08776 [hep-th]].



\bibitem{Brown:1988kg}
J.~D.~Brown and C.~Teitelboim,
``Neutralization of the Cosmological Constant by Membrane Creation,''
Nucl. Phys. B \textbf{297}, 787-836 (1988)


\bibitem{Dvali:2003br}
G.~Dvali and A.~Vilenkin,
``Cosmic attractors and gauge hierarchy,''
Phys. Rev. D \textbf{70}, 063501 (2004)
[arXiv:hep-th/0304043 [hep-th]].


\bibitem{Dvali:2004tma}
G.~Dvali,
``Large hierarchies from attractor vacua,''
Phys. Rev. D \textbf{74}, 025018 (2006)
[arXiv:hep-th/0410286 [hep-th]].


\bibitem{Dvali:2005zk}
G.~Dvali,
``A Vacuum accumulation solution to the strong CP problem,''
Phys. Rev. D \textbf{74}, 025019 (2006)
[arXiv:hep-th/0510053 [hep-th]].



\bibitem{Arkani-Hamed:2006emk}
N.~Arkani-Hamed, L.~Motl, A.~Nicolis and C.~Vafa,
``The String landscape, black holes and gravity as the weakest force,''
JHEP \textbf{06}, 060 (2007)
[arXiv:hep-th/0601001 [hep-th]].


\bibitem{Pisarski:2019upw}
R.~D.~Pisarski and F.~Rennecke,
``Multi-instanton contributions to anomalous quark interactions,''
Phys. Rev. D \textbf{101} (2020) no.11, 114019
[arXiv:1910.14052 [hep-ph]].

\end{thebibliography}
\end{document}